\documentclass[12pt]{article}
\usepackage{amssymb}
\usepackage{amsmath}
\usepackage{amscd}

\begin{document}
\begin{titlepage}
\vspace{1cm}
\begin{center}
{\LARGE Geometry of Axoneme-like Filament Bundles \\}
\end{center}
\vspace{.5cm}
\begin{center}
{\large Andrei Ludu \footnote{E--mail: ludua@nsula.edu}}
\\ \mbox{}  \\
{\it Department of Chemistry and Physics,} \\
{\it Northwestern State University, Natchitoches, LA 71497 USA}
\end{center}
\vspace{.5cm}
\begin{center}
{\large Nathan Hutchings }
\\ \mbox{}  \\
{\it Department of Biology,} \\
{\it Northwestern State University, Natchitoches, LA 71497 USA}
\end{center}
\vskip 1.5 cm

\vfill \eject

\begin{abstract}
We develop a formalism that describes the bending and twisting of
axoneme-like filament bundles. We obtain general formulas to
determine the relative sliding between any arbitrary filaments in a
bundle subjected to unconstrained deformations. Particular examples
for bending, twisting, helical and toroidal shapes, and combinations
of these are discussed. Resulting equations for sliding and
transversal shifting, expressed in terms of the curvature and
torsion of the bundle, are applied to flagellar bend data. We prove
that simultaneous combination of twisting and bending can produce a
drastically drop in the sliding, by decreasing bending rigidity.
\end{abstract}

\vspace{1cm} \noindent {\bf PACS number(s): 87.16.Ka, 05.65.+b,
87.10.+e, 87.17.-d, 02.40.Hw} \vspace*{\fill} \pagebreak
\end{titlepage}
\setcounter{page}{2}

\vfill \eject

\section{Introduction}
\label{sec:1}

Internally generated oscillations of bending and twisting filaments
are of particular interest for the study of cellular biomechanics
and motility in flagellated cells. Flagella and cilia control the
motility of single cell organisms and fluid flow in multicellular
tissues [1]. A current gap in the modeling of flagellar motion is a
mathematical description of the influence of a local deformation on
the global shape (and dynamics) of the filament bundle. Several
groups have addressed deformations [2-14], whereas [15, 16] have
examined the effects of imposed bends on the global curvature sperm
flagella. However, as yet, a means to calculate the precise effects
of a local bend/twist/slide among the individual filaments of the
axoneme on the global geometry of the flagellum has not been
presented. In the present paper, we begin to address that gap by
developing a mathematical formalism based on the differential
geometry of curves, on the theory of motion of curves, and on
functional approaches in order to obtain a new and general formula
for calculating the relative sliding between any two arbitrary
filaments in a bundle of filaments subjected to a 3-d deformation.

To gain a better understanding of how flagellar bends influence
flagellar geometry, we analyze the deformations of axoneme-like
filament bundles [1,2-4,7-15]. Microscopic dynamical models that
describe the occurrence of regular beats, and waves in cilia and
flagella are based on internally driven motion controlled by
molecular motors [2-6,8,10,14,15,17-25]. In these models the
flagellum or axoneme is described as a quasi-continuum
quasi-flexible cylindrical filament bundle containing the
distribution of molecular motors. In the presence of ATP, the motors
generate relative local forces, and local torques, and the resulting
internal stress induces a relative sliding motion of the filaments,
which ultimately leads to the propagation of bending waves
[3,5,17,26]. There is also experimental evidence of the direct
influence of geometrical properties of the filament bundle upon the
dynamics. The are experiments that show how locally induced sliding
produces global bending, in order to conserve the total curvature of
the bundle [16]. Another important motivation for the mathematical
study of the geometry of filament bundles is the intrinsic
regulation of molecular motors through geometry changes induced by
bending [17].

Theoretical models approaching bundle deformations and sliding are
also used to explain symmetry breaking in the nonlinear dynamics of
stiff polymers [28], coupling between internal forces and relative
sliding [5], or analysis of helical structures for modeling bending
and twisting of bacterial flagella, etc.

In this article we introduce the bundle of filaments model and its
deformation, and we obtain in Section~\ref{sec:2} a sliding
expression for bending, in the case of parallel and non-parallel
filaments. In Section~\ref{sec:3}, we introduce the twisting of the
bundle and its influence on the sliding equation. We provide
expressions for cylindrical helix in Section~\ref{sec:31}, and also
for arbitrary helices, and present numerical calculation for a helix
winding around another helix in section~\ref{sec:32}. Examples of
sliding distribution in planar waves, rigid rotation and helical
motion are also discussed. In Section~\ref{sec:4}, we couple bending
and twisting for parallel filaments, and in Section~\ref{sec:5} we
study the most general situation of the sliding produced by
bending/twisting in arbitrary shapes and non-parallel filaments. The
general equation is verified by reducing it to the known 2-d cases
in Section~\ref{sec:52}.

The results in this article describe a mathematical formalism for
predicting the precise position of any filament within a deformed
bundle, which can be used both in numerical or analytical modeling.
This formalism is also useful in experimental studies to predict
local or global distributions of forces and torques (activity of
molecular motors, for example) derived from experimentally observed
shapes. This formalism can be used to better understand the dynamics
of a flagellum, the structure of bends and waves, the importance of
boundary conditions on flagellar shape dynamics, and cellular
propulsion systems that depend on flagella like hair-like systems.

\section{Bending deformations}
\label{sec:2}

The axoneme consists of a tubular structure of 9 pairs (doublets) of
filaments (micro-tubules), and in most, but not all organisms, the
axoneme contains one central pair of filaments which is connected to
the other doublets by protein spokes [1-3,17,26]. The molecular
motors form arms between the outer doublet micro-tubules, resulting
in a local relative sliding displacement between adjacent
micro-tubules [2,5,17]. In addition to a relatively parallel
shifting along the longitudinal axis of the axoneme (providing
micro-tubules are free to move), the boundary conditions imposed on
the doublets creates bending in different directions and twists
around the lateral axis of symmetry. Bending is the result of
holding the doublets together during a shift (quasi-elastic
constrains) produced by the local distribution of axial forces and
normal torques along the flagellum. Local twist results if all the
active pairs slide in the same direction due to torque oriented
along the local flagellum axis, which also generates longitudinal
compression/extension.

Local sliding and twist does not simply correspond with the local
action of motors, since active bending in one region can induce
bending in some passive sections of the bundle. For example, if the
longitudinal displacement of the filaments is restricted at both
ends of flagellum, the total sliding integrated along the flagellum
should be zero. So due to the mechanical constrain of the total
system, regions with positive bending (considered by convention as
positive) should be compensated with regions of negative bending,
even if those regions are not motor active. Very recent experiments
performed with micron calibrated glass needles show if a bend is
forced into an inactive flagellum, the other half bends in opposite
direction trying to keep the total integral curvature zero [15].

In the present model, we study the 3-dimensional deformations of an
axoneme-like structure consisting of a {\it bundle of filaments}. We
define a bundle of filaments $\mathcal{B}$ as a family ${\vec r}_k
(\alpha ): [0,\alpha_{max}]\rightarrow \mathbb{R}_3  )$ of $N$
parameterized regular curves ($k=1,2,\dots, N$) where $\alpha$ is an
arbitrary parameter, common to each curve associated to each
filament. Between any two filaments $k$ and $j$ we can write
\begin{equation}
{\vec r}_k (\alpha )={\vec r}_j (\alpha )+d_{kj} (\alpha ){\vec n}_j
(\alpha)+e_{kj} (\alpha ) {\vec b}_j (\alpha).
\end{equation}
The unit vectors ${\vec n}_j (\alpha )$ and ${\vec b}_j (\alpha )$
are the normal and the binormal directions to filament $j$ at
$\alpha $ [29,30]. The functions $d_{kj}, e_{kj}$ are the {\it
structure parameters} of the bundle, and $k$ runs
$k=1,2,\dots,N$,$k\neq j$. These structure parameters determine the
spatial separation measured in the principal normal plane of
filament $j$, between the reference filament $j$, and any other
filament $k$ in the bundle. If these parameters are constant versus
$\alpha$ for all $j,k = 1,\dots, N$, we refer to this bundle as {\it
parallel}: each two filaments are separated by constant transversal
distances.

We define the {\it relative sliding} of filament $j$ with respect to
filament $k$, at a point ${\vec r}_k (\alpha)$ on filament $k$ by
the expression
$$
\delta_{jk}=[{\vec r}_j (\alpha^{*}(\alpha ))-{\vec r}_k (\alpha)]
\cdot {\vec t}_k (\alpha),
$$
where $\alpha^{*} (\alpha) :[0,\alpha_{max}]\rightarrow
[0,\alpha_{max}]$ is the inverse image $\alpha^* =({\vec
r}_{j})^{-1}$ of the point of intersection between filament $j$ and
the normal plane to the filament $k$ at $\alpha$, chosen such that
the distance between this intersection point ${\vec r}_j (\alpha^*)$
and ${\vec r}_k (\alpha)$ is minimum
$$
\begin{array}{l}
{\vec r}_j (\alpha^* )={\vec r}_k (\alpha)+a_1 {\vec n}_k
(\alpha)+a_2 {\vec b}_k (\alpha ), \ \ \hbox{such that}  \\
||a_1 {\vec n}_k (\alpha)+a_2 {\vec b}_k (\alpha )||_{(a_1,a_2)\in
\mathbb{R}_2} =\hbox{minimum}.
\end{array}
$$
In general, the above intersection equation does not have a unique
solution, because the filament $j$ can intersect the $k$ normal
plane at $\alpha$ in more than one point. That is, we have more than
on pair $(a_1, a_2)$ if filament $j$ bends very much and returns to
this plane. However, in this article we only investigate bundles
where this equation has a unique solution, at any point of every
filament in the bundle. This restricts the bundles with respect to
their curvature and torsion, but these are actually the interesting
real situations. Simply speaking, the relative sliding is the {\it
tangential} displacement between the end points of two close enough
filaments, along their (almost common) tangent. If the filaments are
{\it parallel} and very close to one other, the equations describing
the sliding are accurate, and we analyze such situations in
Sections~\ref{sec:21}, ~\ref{sec:31} and ~\ref{sec:4}. If the
filaments are not parallel, and/or if they are displaced by
separation distances comparable to their length, the above equations
for sliding must account for a more careful geometrical analysis,
which is provided Sections~\ref{sec:22}, ~\ref{sec:32} and
~\ref{sec:5}.

We mention that this general definition of the sliding, i.e.
$\delta_{jk}$ by using the  $\alpha^* (\alpha)$ map defined as
above, is a general one. In the following sections we use this
definition generically, that is we express the sliding for each
particular situation, in some particular convenient coordinates for
that configuration. In that, in the following Eqs.(11,12,19) we use
this definition in a more precise context, by the help of
differential geometry of curves.

A family of more than two {\it parallel} curves are called Bertrand
curves. Their curvature and torsion should be linearly dependent or
any point along the filament, and the only curve that accepts more
than a Bertrand mate is the circular helix (and of course its
degenerated 2-d shape, the circle) [30]. For such an ideal parallel
filament bundle, where there are more than two parallel filaments,
the Bertrand criterium applies, and consequently we can define only
two basic types of ideal deformations: (i) {\it Planar bending},
where the filaments are parallel circles (degenerated helices), and
(ii) {\it Coherent torsion}, where the filaments have helical shape.
Apparently, a relative integrity of the flagellum attachments must
be maintain for normal cell motility in {\it T.brucei} [11], but in
many other real situations the connections between the pairs are
neither constant nor rigid, the filaments are not parallel, and the
general deformations are often intermediate variations of these two
ideal types of shapes.

This situation introduces to relevant considerations. On one hand,
there are real situations when the the shapes are close to circular
bending, to cylindrical helices, or combinations of these, and the
deviation from parallelism of the filaments can be treated as a rare
perturbation. On the other hand, one can analytically treat the most
exact general deformation by assuming that any infinitesimal
deformation is a combination of circular bending and a uniform
torsion, according to the Darboux theorem (an infinitesimal motion
of an arbitrary smooth bundle of curves is an infinitesimal screw
motion) [30]. Thus, one can integrate along the arc length, for
variable radius of curvature, variable plane of bending, variable
pitch (torsion) and variable radius of the helix, and obtain the
precise configuration of the filaments in any shape. The most
interesting experimental application of these considerations is
perhaps when the flagella simultaneously bend and twist [31,32],
with or without variations in the radius of the bundle
[3,6,10,12,31]. None of the current mathematical formalisms for
flagellar shape simultaneously accommodate all of these real-life
(observed) conditions. For completeness, this approach for the
geometry of filament bundles can also be applied to DNA topology
[33], the intracellular cytoskeleton (both radial filaments, and
conical structures) and the mitotic spindle.

\subsection{Circular bending. Parallel filaments}
\label{sec:21}

In this section we describe the (ideal case) of a uniform circular
3-d bending of a bundle of parallel filaments. We consider a number
of identical filaments distributed around a central filament which
is plugged in the origin $O$. The un-deformed bundle, $z<0$, is
parallel to the $Oz$ axis, Fig.1.

A bending of the central filament is defined by a bending center
$O_c$ placed at a distance $R_0 =|{\vec {OO_{c}}} |$ from $O$, at
some angle $\phi_0$ made with the $Ox$ axis, and by the total
bending angle $\alpha_{max}$ which subtends the arc $OO'$ as seen
from $O_c$. The length of the bent region $OO'$ of the central
filament is $L$, and we parametrize
 this part of the filament with
the angle $0 \leq \alpha \leq \alpha_{max}=L / R_0 $. Beyond the
bent region, the central filament is again rectilinear, and it is
confined in the vertical plane making angle $\phi _0$ with $Ox$
axis. The outer filaments (originally parallel to $Oz$ axis for
$z<0$) intersect the $z=0$ plane in points of polar coordinates $(r
cos \phi , r sin \phi ,0)$, namely the initial distribution of outer
filaments in the un-deformed bundle. The outer filaments bend in the
same direction as the central filament, so that all filaments will
uniformly bend in parallel and coaxial arcs of a circle of same
length $L$. Each filament will have a different bending radius $
R(r,\phi )=R_0-r cos(\phi -\phi_0)$, and its own bending center
placed at $x=R_0 \cos\phi_0-r\sin\phi_0 \sin(\phi -\phi_0)$,$y=R_0
\sin\phi_0+r\cos \phi_0 \sin(\phi-\phi_0)$. From these equations we
can obtain the equations of the bent filaments. To calculate the
relative sliding of the filaments, we refer to the {\it normal
terminal surface} of the filaments, intersecting each such curve at
its end point of arc length $L$ (the normal plane of a curve is the
plane perpendicular  to the tangent to the curve) [29]. For small
values of $L$ (shorter arc length, like $\alpha_{max} < 45^o$) the
terminal surface of the filament bundle can be approximated (within
the detectable resolution limits) as a plane. For larger $L$ the
terminal contour distorts into a complex three dimensional
configuration, such as a ``suction cup.''

We define the relative sliding between any two filaments as the arc
length between the two corresponding terminal normal planes,
measured along the filament and having the minimum radius among the
two filaments $ \delta=|\alpha_{max} - \alpha^{'}_{max}| \cdot
\hbox{min} \{ R(r,\phi),R(r',\phi') \} $ which reads
\begin{equation}
\delta=L {{|r \cos (\phi-\phi_0 )-r' \cos (\phi'-\phi_0)|}\over{R_0
    -\hbox{min} \{ r \cos (\phi-\phi_0 ),r' \cos (\phi'-\phi_0) \}}} .
\end{equation}
The radial sliding of a given filament ($\delta_{rad} $) as the arc
length between its end point and its intersection with the terminal
normal plane of the central filament $\delta_{rad}=Lr \cos
(\phi-\phi_0) / R_0$. The relative sliding between any two filaments
placed on the circumference, tangent sliding $\delta_{tan}$, is the
difference between their radial sliding. Similarly, we calculate the
tangent sliding between two parallel adjacent filaments ($r'=r$)
separated with an angle $\delta \phi$ along circumference
\begin{equation}
\delta_{tan}=2Lr sin\biggl ( {{\delta \phi}\over{2}}\biggr )
{{sin({{\delta
        \phi}\over{2}}+\phi-\phi_0)}\over{R_0 -r cos(\delta^{*} \phi +
        \phi-\phi_0)}},
\end{equation}
where $\delta^* \phi =0$ if $0<\Phi-\Phi_0<\pi$ and $\delta^* \phi
=\delta \phi$ if $-\pi <\Phi-\Phi_0<0$. Eq.(3) represents the
tangent sliding between any two filaments in the bundle. To use
Eq.(3), we choose one filament  with angular position $\phi $, and
another one at angular separation $\delta \phi$ from the first one
(clock-wise counting). Then we apply a bending of constant radius
$R_0$, and orientation of bending plane at $\phi _0$, and subtending
an arc length $L$. If we apply Eq.(3) in the range of parameters:
$r=80 nm, \delta\phi=0.7 rad, L=12-20 \mu m$, we obtain values for
the terminal sliding in the range of $\delta_{max}=100-600 nm$ which
is $0.2 - 1.2 \%$ of the total length of the bundle. These numbers
are in agreement with experimental measurements. They provide
sliding in the same range as the sliding obtained from the frequency
and velocity of sliding in [19], the protrusion of doublets measured
with a very good statistics in [7].

As a verification, we notice that Eq.(3) approaches the
2-dimensional bending formula [5,20,23,26,34] in the limit $r<<R_0$
and $\delta \phi \rightarrow 0$
$$
\delta_{tan}  \rightarrow 2Lr  {{\delta \phi}\over{2}}
{{sin({{\delta \phi}\over{2}}+{{\pi}\over{2}})}\over{1 \over
      k}}
\simeq k r \delta \phi \  ds \rightarrow a \int k ds,
$$
where $a=r \delta \phi$ is the linear separation between filaments,
and $k=1/R$ is the curvature of the arc length $L$. The radial
sliding is maximum in the bending plane, while the tangent sliding
has maximum values for those pairs symmetrically placed on the two
sides of the bending plane.

The tangent sliding has maximum value at angles
\begin{equation}
\phi_{\hbox{max sliding}}=\phi_0-{{\delta \phi}\over{2}}\pm
arccos\biggl [ {{r}\over{R_0}}cos\biggl( {{\delta
\phi}\over{2}}\biggr ) \biggr ],
\end{equation}
which is neither in the bending plane nor orthogonal to it. For
example, in the case of the {\it T. brucei} axoneme, by taking
$\delta \phi = 40^o$  and using Eq.(4), the maximum sliding occurs
at about $76^o$ to the right and left of the bending plane, which is
very close to the separation between 3 consecutive pairs. This
result agrees with the experimental evidence of the maximum change
in doublet spacing occurring at doublets 3 and 8, in the geometric
clutch model [2,17]. In a 3-d dynamical approach for curvature and
twist based on the equations of momentum conservation [3] this
effect is explained by the action of the quasi-elastic bridges
between outer doublets. From Eq.(4) we can obtain the maximum
possible bend compatible with a given sliding as a function of the
structural parameters of the axoneme $r, \phi$ and $\delta \phi$. In
the case of {\it T. brucei}, the maximum bend obtained is
$r/R_0\simeq 1.5$ which gives a minimum radius of curvature of
$R_0\simeq 0.9 \mu m$ which is in good agreement with experimental
measurements of curvature [14,15,17,35]. We can also relate the
radius of bending to the maximum sliding $\delta_{tan \ max}$ of the
most active pair with the structural parameters. For a typical
flagellum, where $\delta_{tan \ max}<<L$, we obtain $R_0\simeq r L
\delta \phi / \delta_{tan \ max}$. For a real situation we can
estimate $\delta_{tan \ max}/L=r\delta\phi/R=.03$, which means that
a maximum bend of $R_0=5 \mu m$ can be obtained if the active pairs
slide with as little as 3\% of their total length.

A direct application of the bending equations is shown in Fig.2,
where we calculate the tangent sliding for different motions. We
illustrate the density plot of the tangent sliding for each pair of
filaments in a 9 pair axoneme, versus the arc length. These
(diagonally) oscillating patterns in the sliding could be related
with the activity of motors, showing an interesting self-organized
synchronism. The patterns are in agreement with the "metachronism''
recently obtained in a flagellar model [14,36]. In [13] the authors
noticed patterns similar to Fig.2 (bottom right), for small $L$, and
similar to Fig.2 (top) for short $L$. Calculations in a similar
theoretical model, approaching planar, quasi-planar and helical
waveforms for the sea urchin flagellum resulted in same type of
patterns [16]. From the numerical analysis of the relative tangent
bend in different configurations in Eq.(3), we noticed that the
separation of pairs $\delta \phi$ is not very relevant to the amount
of sliding. The sliding depends stronger on $R$, but after a certain
limit ($R> 15\div 20 r$) $R$ is not any more relevant (asymptotic
behavior).

\subsection{Arbitrary bending. Non-parallel filaments}
\label{sec:22}

We generalize the bending deformation to more general filament
bundle shapes, where all parameters can change along the arc length.
Eq.(2) is valid now only at infinitesimal scale, so we substitute
$L\rightarrow ds$, and integrate the slide from zero to the final
length $L$
\begin{equation}
\delta(L)= \int_{0}^{L} { { r \cos(\phi-\phi_0) -(r+\delta r)
\cos(\phi+\delta \phi-\phi_{0})} \over { R_{0}-\min \{ r \cos(\phi
-\phi_{0}),(r+\delta r) \cos(\phi+\delta \phi -\phi_{0}) \} }} ds.
\end{equation}
where $R_0, \phi_0, r,\delta r,  \delta\phi$ are all functions of
$s$, and $\min$ represent the minimum taken between the two
expressions between $\{ \cdot \}$, for every value of $s$. The
variable geometry of the filament is taken into account by the
curvature $k(s)=1/R_0 (s)$ and torsion $\tau(s)=\partial \theta_0 /
\partial s$ of the filament. The position of each filament is
determined by the polar coordinates $r(s),\phi_0$ and the deviation
from parallelism is related to $\delta r(s) , \delta\phi(s)$. This
equation provides the tangent slide between a filament placed at
$(r, \phi)$ angular coordinate in the bottom circumference, and
another filament at coordinates $r+\delta r, \phi+\delta \phi$
(separated with $\delta \phi$ from the first one). During this bend
the radius of curvature $R_0$ and the orientation of the bending
plane $\phi_0$ can change, and the arc length of the bundle is $L$.
All these parameters are variable versus $s$, so there is no
parallel constraint, and the bend is no longer circular and uniform.
However, because the twisting is not yet taken into consideration in
Eq.(5), there are some restrictions on the shapes. For example, if
we keep $R_) (s)$ constant, and have the bending plane rotate
uniformly $\phi_0 (s)\sim s$ we do not obtain a cylindrical helix,
but a bend helix twisting around a toroidal shape. The full
generalization of sliding and twisting, for any shape, will be
discussed in Section~\ref{sec:4}.

In the following, we discuss the consequences of deviation from
parallelism upon the sliding. Such deviations can be produced by
tangential deviance of filaments along the circumference controlled
by $\delta \phi $, or by radial deviance of filaments controlled by
$r(s)$. We introduce a measurement of the tangent and the radial
deviations from parallelism by
\begin{equation}
\nshortparallel _{tan}(s)={{r (\delta \phi (s)-\delta \phi
(0))}\over {s}}, \ \nshortparallel _{rad}(s)={{r(s)-r(0)}\over {s}},
\end{equation}
respectively. Basically $\nshortparallel _{tan}=d_{kj}$ and
$\nshortparallel _{rad}=e_{kj}$ from Eq.(1). The change in the
sliding produce by these deviations will be measured by $\triangle
\delta (s) =1-\delta (L,\phi )|_{ \delta \phi (0), r(0),\dots )} /
\delta (L,\phi )|_{(\delta \phi (s), r(s),\dots )}$. The analytical
expression of $\triangle \delta $ can be obtained from Eq.(2)
through the theorem of derivation of implicit functions, but we do
not go into such details. Instead, we present some numerical
evaluation of the sliding for arbitrary shapes, Eq.(5), for
different deviations in parallelism. In Fig.3a we illustrate the
sliding of 5 parallel pairs having different positions with respect
to the bending plane ($\phi=0, \pm 30^o$, and $\pm 60^o$) in a
cylindrical bundle. For a circular bend, for example, the sliding
uniformly increases in magnitude with $L$, accordingly to Eq.(3).
Then we choose different deviations from parallel and represent in
Fig.3 the new sliding, for the same bend by a dotted line for each
pair correspondingly (using now Eq.(5). In some of the frames we
also represent the local infinitesimal slide, i.e. the integrand of
Eq.(5) by the thin oscillating lines. These data clearly indicate
that minor perturbations in parallelism have only small effects on
the overall sliding of the filaments.

For tangential deviation from parallelism given by an oscillatory
function $\delta \phi (s) \sim \delta \phi_0 \sin (\omega s ) $,
with $\delta \phi_0 =0.07 rad$ taken from the clutch model in [10],
there is little total change in sliding $\pm 5 nm$ (Fig.3 upper left
frame). But, if the tangential deviation of parallelism is
constantly increasing with the arc length (Fig.3b) the sliding is
more enhanced. The same behavior occurs for radial non-parallelism.
If the bundle radius oscillates (Fig.3c) the total new sliding
follows the radius changes but the effect is still small. If we
overlap the two types of deviations from parallelism, the changes in
the total sliding is more significant (Fig.3d).

In this case we can have (for appropriate resonance between radial
and tangent oscillations like $\delta r=\pm 2-5 nm$ and $\delta\phi
(s)-\delta \phi(0) =5^o$ [12]) a change in sliding of more than
200\% over the total length of the bundle. In Fig.4 we show a
variable radius, variable pitch bend (the shape is sketched in the
left upper corner of the frame. Such situations are discussed in
[12] where the authors investigate uniform ($3.5\mu m$ for each
doublet) and non-uniform ($10\mu m$ per one doublet) tangent shift
of the doublets. The thin continuous lines represent sliding of the
same 5 pairs for parallel bundle. If we consider simultaneously the
two types of non-parallelism, as oscillations in radius and in
angular separation for each pair, we note that this "intertwining"
of filaments stabilizes and lowers the total sliding (dotted
curves). The same effect is present if one considers the twisting of
filament, as we will show in Section~\ref{sec:5}. Finally, in Fig.5
we present the sliding between three adjacent filaments, where the
outer filaments are parallel and the middle filament-2 moves along
$s$ from closer to filament-1 at one end (left cross section) to
closer to filament-3 at the other end (right cross section). This
case is most critical to the geometric clutch model [10,17]. We
choose a simple circular bend and plotted the sliding for parallel
filaments ($\delta_{12},\delta_{23}$) (continuous curves) and the
sliding for tangentially oscillating filaments (dotted curves,
($\delta_{12\nshortparallel },\delta_{23\nshortparallel }$)). Of
course, by using Eqs.(2,3), we can approach more complicated
distortions of the circular symmetry, like the splitting patterns
observed  for eukaryotic flagella and cilia axoneme [9].

\section{Twisting}
\label{sec:3}

Long flexible structures like flagella and cilia are subject to
twisting deformations in addition to bending as evidenced by both
planar and helical bending patterns within the same organism under
different conditions [3,6]. Internally driven bending is produced by
a distribution of torques always oriented along the normal to the
bundle envelope. Bending deformation may be  the action of any
number of pairs, while twisting deformation requires simultaneous
action of two (or any even number of) pairs, since all pairs have to
slide in the same direction. According to the "geometric clutch
model" [10,17] and [35], it is unlikely that the axoneme is
perfectly cylindrical along its entire length, and according to
[12], the diameter of the axoneme can also oscillate. In this
section, as a starting point, we calculate the filament twist in the
cylindrical approximation. In the case of small changes in the
bundle radius, or small radius oscillations, we can still use this
cylindrical approximation, and treat the extra sliding produce by
deviation from the cylindrical shapes, as perturbations. For example
for a 3-d sliding model, if the twist is produced by fixed links
located asymmetrically, the resulting sliding is a second-order
effect, proportional to the square of the flagellum radius [2,3].
Although the approximation is reasonable, in Section~\ref{sec:4} we
unconstrain this cylindrical-parallel approximation, and we
calculate the sliding produced by twisting for general shapes.

\subsection{Uniform twisting. Cylindrical helix}
\label{sec:31}

We consider a circular bundle of radius $r$ consisting of N
equidistant parallel pairs of length $L$. For each pair, the two
parallel filaments ($aa'$ and $bb'$) are separated by an angle
$\delta \phi$, Fig.6. In a twisting deformation we assume that the
upper ends $a'$ and $b'$ of the filaments are rotated with the angle
$\phi$ (which is also the parameter along the filaments), while
their basis remain fixed. The height of the bundle decreases from
$L$ to $h$. The filaments take the shape of two parallel cylindrical
helices of radii $r$, height $h$, and pitch $b$, where the pitch is
defined as $h=b\phi_{max}$. By using the formula for the length of a
helix $L=\sqrt{r^2 +b ^2} \phi$, we obtain the twisting slide in the
form
\begin{equation}
\delta_{twist}={{a r \phi}\over{L}}={{r^2 \delta
\phi}\over{\sqrt{r^2+b^2}}},
\end{equation}
where $a=r \phi$ is the linear separation between filaments in a
pair. This twisting sliding is quadratic with respect to filament
separation, and hence Eq.(7) is a good working approximation for
situations where the deviation from parallelism are insignificant.
The relative sliding for a bend is larger than the sliding for a
twist for the same amount of work. That is, twisting may act like a
lower gear, while bending may act like a higher gear in terms of
self-propulsion. The relative shift between filaments during a twist
is constant along the arc length, so shift does not increase versus
length like in the case of the bending deformation.

Furthermore, twist results in compression of total length of the
filament. The compression of the total length of the bundle for a
given twist is $\delta L=L-\sqrt{L^2-r^2 \phi^2}$ and the relative
compression can be approximated with $ {{\delta L}\over{L}}\simeq
{{r^2 \phi ^2}\over{2 L^2}}={{r k}\over 2}$. For example, for a
bundle of radius $r=1\mu$m, length $L=10\mu$m, twisted with a
complete turn $\phi=2\pi$, with pairs separated at $\delta \phi
=5^o$ the shift between the adjacent filaments is approximately $100
nm$ and the relative length compression is $\delta L/L=1\% $. We can
test the twisting slide formula, Eq.(7). Since a uniform helix is
produced by an infinitesimal bending at the base circle, by consider
Eq.(3) with $R=a,L=a \cos \alpha$ we obtain $ \delta_{bend}
|_{infinitezimal}=L a / R_0 =a \cos ( \pi / 2 -\alpha )
=\delta_{twist}$. Within any 3-d deformation of a dynamic filament
bundle, some twist among the filaments will occur. To date, the
influence of filament twisting has been neglected, possibly because
the twist will often result ahead of or behind a region of the
bending where the attention has been focused. A numerical analysis
of twist generated in an active region and accumulated along the
length of the axoneme is presented in [4]. For an axoneme of radius
$r=80 nm$, linear separation between doublets $a=40 nm$, length
$L=20\mu m$ and total twist $\phi =0.2$ rad, Eq. (7) results in a
sliding $\delta_{twist}=30 nm$ which is in good agreement with other
numerical models, and experiments [4]. In this section we considered
that filaments remain parallel while twisting, and consequently are
confined in a cylindric surface. This simplifying hypothesis is
revised in Section~\ref{sec:3} and Eq.(7) is substituted with a
formula valid for twisting around variable geometry bundle. However,
even if the filaments begin to taper, splay, or intertwine, [15,35],
hence loose parallelism, Eq.(7) can still give a close approximate
evaluation of the situation.

\subsection{Generalized twisting. Arbitrary helix}
\label{sec:32}

The bundle shape may change in time in response to the collective
internal and external forces, such as in the case of {\it T.brucei}
where the flagellum is wound around the cell as a variable
radius-variable pitch helix [1,23,26]. In the following, we extend
Eq.(7) for uniform sliding of a cylindrical helix to general
twisting deformation when the filament bundle is itself deformed.
Let $\Gamma$ curve be the central filament of this bundle described
by an equation ${\vec  r}_{\Gamma} (\alpha)$ having its
Serret-Frenet trihedron given by ${\vec T}(\alpha)$ (the unit
tangent vector), ${\vec N}(\alpha)$ (the principal normal vector)
and ${\vec B}(\alpha)$ (the binormal vector), and metrics
$G(\alpha)=\partial {\vec r}_{\Gamma} /
\partial \alpha \cdot
\partial {\vec r}_{\Gamma} / \partial \alpha $ [30].
In order to twist the filaments around the central filament we
construct, for every $\alpha$, a circle of radius $r(\alpha)$
centered in ${\vec r}_{\Gamma}(\alpha )$, and parallel to the
principal normal plane of $\Gamma$ at $\alpha$. Each such circle is
parameterized by a new variable $\phi_p$ measuring the local twist
of the pair at $\alpha $, around the local $\Gamma$ axis. The
surface generated by this family of circles is ${\vec c} (\alpha,
\phi_p )={\vec r}_{\Gamma}(\alpha)+ r ({\vec
  N}(\alpha) sin \phi_p  -({\vec B}(\alpha) cos \phi _p )$.
In order to construct such a variable helix, we have to relate the
$\alpha $ and $\phi_p$ parameters. The local (infinitesimal) height
along this variable helix is equal to the corresponding
infinitesimal element of the length of the supporting curve
$\Gamma$, that is $ds_{\Gamma }=b d\phi_p$ where $b$ is the local
pitch of this infinitesimal helix. Then, the variable helix $\gamma$
makes a constant angle with the tangent to the central filament
curve $\Gamma$, namely $\tan \psi =b$. It results in the following
relation between $\phi _p$ and $\alpha$: $ \phi _p (\alpha)={1 \over
b}\int_{0}^{\alpha}G^{1 \over 2} (\alpha') d\alpha ' $, which
provides the metrics of the variable helix
\begin{equation}
g(\alpha)=G(\alpha ) \biggl [ \biggl ( 1-r k_{\Gamma} \sin ( \phi_p
(\alpha) ) \biggr )^2 +r^2 \biggl ( \tau _{\Gamma} +{1 \over b}
\biggr ) \biggr ].
\end{equation}
Here $k_{\Gamma}$ and $\tau_{\Gamma}$ are the curvature and the
torsion of the $\Gamma$ central filament, respectively. The
parameter $\alpha $ of the $\Gamma$ curve measures the deformation
of the whole bundle, and its maximum value at the end of the bundle
is $\phi_b$. All filaments in the bundle are described by the
metrics in Eq.(8), the difference being made by the initial angular
shift in $\alpha$ at the beginning of the bundle. From Eq.(8) one
can obtain the curvature and torsion of any filament ($\gamma$
curve), and then calculate its length at a given $\alpha$. For any
two such filaments the difference of their lengths at $\alpha =
\phi_b$ is their relative sliding.

The above procedure can calculate the twisting slide for any shape,
but in the following we present an explanatory example where the
bundle has a helical shape of radius $R$ and pitch $B$. The
filaments lye on the external surface of the bundle. Locally, the
central filament $\gamma$ is an infinitesimal helix having its
principal normal always parallel to the normal plane to the bundle
surface. Consequently the curvature of the central filament,
$k_{\gamma}$, is equal to the normal curvature of the bundle
surface, and consequently is bounded between the values of the
principal curvatures $k_{\Gamma 1,2}=\{ 1/r,r/(R^2+B^2) \}$ of the
bundle surface [29]. The curvature $k_{\gamma}$ can be approximated
with
\begin{equation}
k_{\gamma} \simeq \left\{\begin{array}{rcl}
k_{\Gamma}={{R}\over{R^2+B^2}} & \mbox{if} & b>b_{lim} \\
k_{\gamma 0}={{r}\over{r^2+b^2}} & \mbox{if} & b<b_{lim},
\end{array}\right.
\end{equation}
where $b_{lim}=\sqrt{{{(R^2+B^2) r-R r^2}\over{R}}}$ is the critical
value of the pitch of the filament. The relative sliding of the
filaments can be calculated with good approximation by using
Eqs.(3,8,9) by using again the fact that a local twist is equivalent
to an infinitesimal bending. We denote the radius and the pitch of
the bundle helix with $R,B$. The bundle is a cylinder of radius $R$
twisted as a helix with total length $L_b=\sqrt{R^2+B ^2} \phi_b$,
where $\phi_b$ is the total twist angle of the bundle. We substitute
in Eq.(3) $L\rightarrow \sqrt{g(\alpha)} d\alpha$ and $\phi_0
\rightarrow \alpha$, with $g(\alpha)$ from Eq.(8). With this
notations, we can find the sliding within the $\gamma$ pair by using
Eq.(9).
\begin{equation}
\delta_{\gamma} \simeq \delta \phi \cos \biggl ({{\delta
\phi}\over{2}}+\phi-\alpha \biggr )  \left\{\begin{array}{rcl}
{{r^2R}\over{b}}\sqrt{{{R^2+B^2}\over{r^2+b^2}}} & \mbox{if}
& {{R^2+4B^2-{{r R}\over 2}}\over {r\sqrt{R^2+3B^2}}}>1, \ \ {{(R^2+B^2-Rr)r}\over {b^2R}}>1 \\
{{r R}\over{b}}\sqrt{{{r^2+b^2}\over{R^2+B^2}}} & \mbox{if} &
{{R^2+4B^2-{{r R}\over 2}}\over {r\sqrt{R^2+3B^2}}}>1, \ \
{{(R^2+B^2-Rr)r}\over {b^2R}}<1 \\
{{r R}\over{b}}\sqrt{{{r^2+b^2}\over{R^2+B^2}}}  & \mbox{if} &
{{R^2+4B^2-{{r R}\over 2}}\over {r\sqrt{R^2+3B^2}}}<1,
\end{array}\right.
\end{equation}
where $\delta \phi$ is the angular separation between the filaments
in the pair, and $\phi$ is the angular coordinate of the pair around
the circumference.

For example we take  $L=12 \mu$ length of the flagellum, $r=0.5\mu
$, a separation between pairs of $\delta \phi ={{\pi}\over 9}$, and
a full twist of $2 \pi$ around the cell body (helices with variable
radius and total bending plus twisting sliding will be calculated in
the next section), like in the case of {\it T. brucei}). In the case
of very twisted cell (auger shape) we choose $b={{12}\over {2 \pi}}
\mu$, $R=1 \mu$, and $B={{12}\over {2 \pi}} \mu$. The pitches $b,B$
are calculated from the helix equation $Height_{helix}=b
\alpha_{max}$. These parameters place us in the second row of
formula Eq.(10). The maximum twisting sliding in this shape is
$\delta \geq 7 nm$ or $\delta /L \geq 0.5 - 0.7 \%$.

In the case of an elongated cell body, we have $b={{12}\over {2
\pi}} \mu, R=6 \mu$, and $B={{6}\over {2 \pi}} \mu$ and consequently
we have a maximum slide of $\delta /L=1 \%$. The main sliding
control parameters are $R/r$ and $B/R$. These results are in good
agreement with Eq.(7) for a cylindrical helix with uniform twisting.
Even for a very small curvature and extreme bending, $R/r<3/2$, we
can still apply the same Eq.(12). Such estimations show that
strongly coupled bending and twisting produces small sliding effects
(percentages of the bundle length), and this sliding is not strongly
dependent on the variable curvature (on the bending radius). Thus,
in a swimming stroke, the terminal sliding oscillates between $.5\%$
and $3\%$ which is still small compared to a bending without twist
for the same length. An optimal combination between sliding and
twisting could adapt the swimming regime of the cell to the external
viscosity, by changing the pattern from helical to planar geometry
[6,7].

From numerical calculations with different values for $r,b,R,B$ we
noticed that the curvature of the flagellum is oscillating around
the largest of $k_{\Gamma}={{R}\over{R^2+B^2}}$ and $k_{\gamma
0}={{r}\over{r^2+b^2}}$, that is the the bundle curvature, and the
curvature of the same filament as a cylindrical helix. A filament
has a maximum curvature when $k_{\Gamma}\simeq k_{\gamma 0}$, i.e.
when there is some geometric resonance between the two helices.
Consequently, the average sliding almost doubles in such a
situation. When the geometric parameters that mimic the shape of a
{\it T.brucei} go from a relaxed (elongated) position
($R=10\mu,r=1.5\mu,B=30/2\pi \mu,b=12/2\pi \mu$) towards an auger
twisted position ($R=5\mu,r=1.5\mu,B=6/2\pi \mu,b=12/2\pi \mu$) the
average sliding almost doubles, from $4\%$ to about $8\%$, which may
relate to the mechanism by which these cells can effectively migrate
through host tissue.

\section{Mixed deformations. Parallel filaments}
\label{sec:4}

Herein we introduce a general sliding formula for simultaneous
bending and twisting in 3-dimensions. We construct the bundle
starting from its central filament ${\vec r}_C (s)$ ($s$ is the arc
length) and constructing a smooth family of circles of radius $\rho
=\rho (s)$ all centered in ${\vec r}_C (s)$ and lying in the normal
plane of the central filament at $s$. Each outer filament $i$ is
described by a curve ${\vec r}_{i}(s_{i})$, $i=1,2,\dots , N$,
intersecting each circle only once. No outer filament can return
towards the initial point, that is ${\vec t}_C (s) \cdot {\vec t}_i
(s)>0$ for any $s$ and $i$. We us choose two arbitrary filaments (1
and 2) initially separated by $a$ at $s_1 =s_2 =0$, and calculate
their relative sliding at $s=L$, $\delta (L)$. We also assume
$a,\delta (L)<<L$, Fig.7.

The normal plane of filament 1 at $s_1 =L$ (defined by ${\vec n}_1
(L)$ and ${\vec b}_1 (L)$) intersects filament 2 at some $s_2=s^*
\neq L$ producing a sliding along the tangent direction
$\delta(L)=L-s^*$. Under the hypothesis of small separation of pairs
compared to the length of the bundle we can expand ${\vec r}_2 (s^*
)$ in Taylor series around $s_2=L$. The intersection condition
between the normal plane and filament 2 becomes ${\vec n}_1 (L) C_1
+{\vec b}_1 (L) C_2 +{\vec r}_1 (L)-{\vec r}_2 (L) =-{\vec t}_2 (L)
\delta $, where $C_{1,2}$ are the coordinates of the intersection
point in a local 2-d frame in the normal plane. In the approximation
${\vec t}_1 (L) \simeq {\vec t}_2 (L)$ the infinitesimal sliding
becomes
\begin{equation}
\delta (L) = \Omega|_{s,a} ({\vec r}_2 -{\vec r}_1 ) _{s=L} \cdot
{\vec t}_1 (L),
\end{equation}
where $\Omega$ is the antisymmetric part of the second order
differential with respect to $s$ and $a$ defined as $\Omega
|_{s,a}=(1+s{{\partial}\over{\partial s}}|_{s=a=0}
+a{{\partial}\over{\partial a}}|_{s=a=0} +
as{{\partial^2}\over{\partial s\partial a}}|_{s=a=0})$. Technically,
$\Omega$ is provided by the second-order antisymmetric terms in the
Taylor expansion of Eq.(11) with respect to $s$ and $a$ around
$(0,0)$. We need to take into account the second differential
because the sliding produced by twisting is on order of magnitude
smaller than the sliding produced by bending. The total slide is
obtained by integration of Eq.(11) along $s\in [0,L]$. The operator
$\Omega $ is the second order prolongation of the infinitesimal
translation generator in $s$ and $a$ acting on the filament bundle
surface, [30], and the infinitesimal sliding is the local action of
this operator on the $({\vec r}_1 -{\vec r}_2 ) \cdot {\vec t}$
function. This model can be tested by some simple examples. In the
case of a cylindrical helix, the sliding in Eq.(11) is identical
with Eq.(7) for twisting. Also, for a circular bend, if we put
$a=\delta \phi R$ and $\phi -\phi_0 =\pi/2$, Eq.(11) approaches
Eq.(3) for bending in the limit $\delta \phi_0 \simeq sin (\delta
\phi_0)$.

\section{Bending and twisting deformations. Non-parallel filaments}
\label{sec:5}

In this section we present a formula for the total sliding produced
by simultaneous bending and sliding, for any type of geometry with
all parameters variable. Of course such a formula accommodates
non-parallel filament configurations, and approaches in the limiting
situations all the results presented in the previous sections. We
describe the filament bundle as a family of smooth (at least second
order differentiable) curves ${\vec r}(s_{\beta}, \beta)$ each
parameterized by its the arc length $s_{\beta}$ and by $\beta$ to
label the family. For each such curve we describe its metrics
$g(\alpha, \beta)$ and the infinitesimal arc length $ds=\sqrt(g)
d\alpha$. There are two possible interpretation of this last
parameter; (1) $\beta$ can describe the deformation of a certain
filament, and hence $\beta$ can be related to the time; or (2)
$\beta$ can describe the mapping of a (deformed) filament into its
neighbor pair filament, so it would be related to separation and
non-parallelism.

\subsection{General sliding formula}
\label{sec:51}

For a given point $\alpha, \beta$ along one filament its
infinitesimal displacement in both parameters variations is
\begin{equation}
{\vec dr}={\vec t}\sqrt{g}d\alpha +({\vec t}\delta + {\vec
n}\Delta+{\vec b} \Lambda)d\beta ,
\end{equation}
where $\{ {\vec t}, {\vec n},{\vec b} \}$ is the local Serret-Frenet
trihedron of the filament. The first term in the RHS (proportional
with displacement $d \alpha$ along each filament) describes the
regular advancement of the point along the curve by increasing the
arc length $s$. The three shifting functions $\delta
(\alpha,\beta)$, $\Delta(\alpha,\beta)$ and $\Lambda(\alpha,\beta)$
characterize the displacement of the position of a point, placed at
a certain fixed distance $s(\alpha)$ from the base ($\alpha=0$),
when we go from one filament to another (modify $\beta$). They
represent shifting of points as follows: $\delta$ is the shift along
the tangent of the filament (sliding), so when we change $\beta$,
and move from one filament to another this function measures the
actual sliding at any point $\alpha$ of the filament during this
transformation. The function $\Delta$ measures the shift along the
principal normal of the filament, that is the separation of
filaments in the bending plane. The last function $\Lambda$ measures
the shift along the binormal, that is the relative displacement of
filaments perpendicular on the bending plane and on the tangent to
the filament. For example, in a planar bending we have from Eq.(1)
the interpretation of the functions in Eq.(12) in terms of
separation and deviation from parallelism, $\Delta= r\delta \phi +
\nshortparallel_{tan} $ (also for a cylindrical helix) and
$\Lambda=\nshortparallel_{rad}$.

From the smoothness property of the mathematical curves that
describe the filaments we can use the symmetry of second order
derivatives and obtain the dynamical equation for the metrics [37]
\begin{equation}
{{\partial g}\over{\partial \beta}}=2\biggl ( \sqrt{g} {{\partial
\delta}\over{}\partial \alpha}- g k \Delta\biggr ) .
\end{equation}
By differentiation with respect to $\beta$ the equation for the arc
length is $s(\alpha,\beta)=\int_{0}^{\alpha}\sqrt{g}d\alpha'$, and
by using Eq.(13) we obtain
\begin{equation}
{{\partial L(\alpha_{max},\beta)}\over{\partial s}}=\delta
(\alpha_{max}, \beta)-\delta (0,
\beta)-\int_{0}^{\alpha_{max}}k\Delta \sqrt{g} d\alpha.
\end{equation}
The deformations, bending and twisting, change just the shape of the
filaments, and not the total length of each filament, so we have the
local conservation of the length with respect to deformations
$({{\partial s}\over{\partial s}})$. If we consider no sliding at
the beginning of the bundle ($\delta (\alpha=0) $), we obtain for
sliding the expression
\begin{equation}
\delta (\alpha,\beta)=\int_{0}^{\alpha} \sqrt{g}k\Delta
d\alpha'=\int k {\vec {dr}}\cdot {\vec n} ds.
\end{equation}
The full description of  the 3-d shifting $\delta,\Delta,\Lambda$
can be related to the geometry of the filament (curve completely
described by the initial position and by the curvature
$k(\alpha,\beta)$ and torsion $\tau(\alpha,\beta)$, [29], through
two partial integro-differential equations. Following the theory of
curve motion [37] we have
\begin{eqnarray}
{{\partial k}\over{\partial \beta}}= {{\partial^2
\Delta}\over{\partial s^2}}+ (k^2-\tau^2) \Delta+ {{\partial
k}\over{\partial s}} \int_{0}^{s} k \Delta ds' - 2 \tau {{\partial
\Lambda}\over{\partial s}}-\Lambda {{\partial \tau}\over{\partial
s}}   \nonumber\\
{{\partial \tau}\over{\partial \beta}}={{\partial }\over{\partial
s}} \biggl [ {1\over k} {{\partial }\over{\partial s}} \biggl (
{{\partial \Lambda}\over{\partial s}}+\tau \Delta \biggr )
+{{\tau}\over {k}}\biggl ( {{\partial \Delta}\over{\partial s}}-\tau
\Lambda \biggr ) +\tau \int_{0}^{s}k\Delta ds' \biggr ] +k\tau
\Delta+ k{{\partial \Lambda}\over{\partial s}}
\end{eqnarray}
These equations are written in terms of the arc length $s$, but it
is easy to substitute it with $\alpha$ through the transformation
$ds=\sqrt{g(\alpha, beta))}d\alpha$. With Eqs.(15,16) we can solve
the geometry (or kinematics, if $\beta$ is time) problem of the
filament bundle. The {\it direct problem} consists in finding the
shape of each filament in the bundle if we are given the full set of
three shifts: tangent and two transverse. Namely the functions
$\delta,\Delta,\Lambda$ depending on $\alpha$, the evolution along
the filament, and $\beta$, for each filament. For a 9+1 axoneme for
example, $\alpha$ takes values between zero at the basal plane, and
$\alpha_{max}, s(\alpha,\beta)=L$ for a prescribed length $L$, and
$\beta=1,2,\dots,10$ is integer number. Consequently, for given
functions $\delta,\Delta,\Lambda$, Eqs.(15,16) can be integrated
with respect to the unknown functions $g(\alpha,\beta),k(\alpha,
\beta), \tau (\alpha,\beta)$, within the prescribed initial data.
Next, knowing the metrics, curvature and torsion we can integrate
the fundamental equations of the differential geometry of curves
[29,30], and obtain the curves ${\vec r}(\alpha,\beta)$, that is the
shapes of any filament in the bundle. Conversely, the same
Eq.(15,16) can solve the {\it inverse} problem, that is: given the
shape of all filaments (the curves and consequently their metrics,
curvature and torsion) we can integrate the equations and obtain the
distribution of all three shifts for any filament in any point. Such
an example is illustrated in Fig.8.

Both directions of integration are possible (though tedious
numerical calculations may be involved) because the dynamical system
described by Eq.(16) (having as parameter $\beta$ instead of time)
is an integrable system belonging to the MKdV- or NLS-equation
hierarchies [37]. Eqs.(15,16) allow us to find the shape compatible
with any distribution of sliding and twisting, and conversely, for a
given shape to obtain the distribution of sliding and twisting. If
one can relate the motor activity with all the 3-d sliding and
shifts in a one-to-one correspondence, this approach could help a
better understanding of the dynamics of filament bundles.

To illustrate with a simple, we choose a slightly deformed
cylindrical helix, such that the curvature and torsion can still be
considered $s$-independent, having almost constant bundle radius
($\Delta \simeq 0$), but with filaments free to change their angular
distribution in the cross section (non-parallel). We have $\Lambda
\simeq \delta$, and from Eqs.(16) it results that ${{\partial^2
\Lambda}\over{\partial s^2}}=0$, and hence
$\Lambda(s,\beta)=\Lambda_{0}(\beta) s+\Lambda_{1}(\beta)$. This
means that for any filament in the bundle (any $\beta$) the shift
along the helical axis (along the binormal unit vector ${\vec b}$)
is constantly increasin/decreasing with a ratio depending from
filament to filament. This linear variable shift is proportional to
the pitch of the helix. Eqs.(15,16) are nonlinear in all variables,
but if we consider unknown only $\delta, \Delta,\Lambda$, or
conversely in $g,k,\tau$, they become linear. This partial linearity
allows the evolution of geometry to follow the evolution of
shifting/sliding, and the other way around. Since these linear
equations are of order three we also expect high dispersion, which
means that localized zones of high sliding will disperse and
re-distribute along $s$ in time or from filament to filament. For
example, an oscillating distribution of sliding and transverse
shifting will produce same oscillations in curvature and torsion,
generating waves and helices. On the other hand, such self-organized
oscillations of internally shifted filaments were described in
previous 2-dimensional models [5]. Wavelike propagating shapes were
obtained by modeling a Hamiltonian system, and special behavior
(bifurcations, spontaneous oscillations) was identified as critical
for the self-generation of wave patterns [20,25,28,31]. Such type of
behavior, even presence of solitons, can be already predicted from
the nonlinear structure of Eqs.(16), as we describe further in
section~\ref{sec:52}, without any reference to any particular
physical model.

In the following, we focus on the sliding eq.(15) and we calculate
the relative separation vector ${\vec dr}$ between the ends of two
given filaments ($i,j$), that is ${\vec dr}(s) ={\vec r}_i (s)
-{\vec r}_j (s)={\vec r}_{ij}(s)$, ${\vec
  r}_{ij}(0)={\vec r}_{ij0}$, $|{\vec r}_{ij0}|=a$, where $a$ is the
initial linear separation between these 2 filaments. During the
displacement $ds$ of the $i$ filament the vector ${\vec r}_i$
performs an infinitesimal rotation of angle $d\phi=k ds$ around the
binormal (local bending), and a rotation of angle $d\psi=\tau ds$
around the tangent (local twist). Thus, its variation reads $d{\vec
r}_i=({\vec b}\times {\vec r}_i ) d\phi +({\vec t}\times {\vec
  r}_i ) d\psi$. The differential equation governing ${\vec r}_i$ vector is
\begin{equation}
{{d {\vec r}_i}\over{ds}}={\vec \omega} \times {\vec r}_i , \ \ \
{\vec \omega}=k {\vec b} +\tau {\vec t}.
\end{equation}
with initial conditions ${\vec r}_i(0)={\vec r}_{i0}$,  namely the
structural position of the filament $i$ in the initial cross section
of the un-deformed bundle (the base circle). Eq.(16) represents a
rotation which copies the Darboux motion of the normal unit vector,
since we have $d {\vec n} / ds={\vec \omega} \times {\vec n}$, with
the exception that the origin of the vector ${\vec r}$ also
translates along the central filament with the arc length $s$ [30].
Eq.(16) is integrable and its solution has the form
\begin{equation}
{\vec r}_i (s)=e^{\int_{0}^{s}{\hat \omega} ds'}{\vec r}_{i0}.
\end{equation}
where the exponential is defined in the exponential matrix operator,
and ${\hat \omega}$ is the dual tensor of the vector ${\vec
  \omega}$,
that is ${\hat \omega}_{nm}=\epsilon _{nmp}{\vec  \omega}_p$ with
$n,m,p=1,...3$ and $\epsilon _{nmp}$ being the Levi-Civitta tensor.
By using Eqs.(15,17) we obtain the most general sliding formula for
a a pair (i,j) of filaments in a bundle described by the normal
${\vec n}$ and curvature $k$
\begin{equation}
\delta_{ij} (s)=({\vec r}_{i0} -{\vec r}_{j0})\cdot \int_{0}^{s}
e^{\int_{0}^{s'}{\hat \omega} ds''} \cdot {\vec n}_i (s') k_i (s')
ds'.
\end{equation}
We mention that there is no equivalent equation for general 3-d
sliding in the literature, so we can compare Eq.(18) only with
numerical results. As a general check, if we choose a plane curve
(zero torsion and constant binormal vector) the integral $\int {\hat
\omega } ds$ reduces to a rotation of angle $\theta =\int k ds$
about the Oz axis. The right-action of this matrix through the
second dot product rotates the normal ${\vec n}(s)$ back to its
initial direction ${\vec n}(0)$, so we have $({\vec r}_{i0} -{\vec
r}_{j0})\cdot {\vec n}(0)=\Delta = a$, as in the 2-d sliding
formulas from literature.

We can use Eq.(18) to calculate and predict the relative sliding for
different pairs in different configuration of simultaneous bending
and twisting. We performed several numerical checks of consistency
between Eq.(18) and the other expressions obtained in this article
for the sliding, namely Eqs.(2,3,5,7,10,11), and we found a very
good match for different geometries. For example a helix of
curvature $k={{R}\over {R^2+b^2}}$ and torsion $\tau ={{b}\over
{R^2+b^2}}$ has ${\vec \omega}=(0,0,g^{-{1\over 2}})$, so the only
nonzero elements in the matrix are ${\hat
  \omega}_{12}=-{\hat \omega}_{21}=-g^{-{1\over 2}}$, and the exponential
of this matrix is a rotation matrix around the Oz axis. By using
this rotation matrix, and a particular solution of Eq.(17) for a
bundle of nine pairs, we can calculate from Eq.(18) the sliding at
any point $s$ of the bundle and for any shape described by the
corresponding curvature $k$ and normal vector ${\vec n}$.

In general, this sliding analysis provides local and global
information about the geometric constraints of the bundle. For
example, if the total slide at the end of a given pair is zero, the
global integral curvature of this curve should also be zero. Thus,
sliding directly interacts with the dynamics.

For example, certain boundary conditions at the end of the pairs
(like clamped ends) introduce restrictions in the class of
admissible shapes. An interesting situation (which has not been
mentioned in literature) occurs in the mixed bending and twisting
case. For pure bending the relative sliding of a pair in a bundle is
constantly increasing versus the length of the bent segment (also
proportional to the bending angle). If we first twist the
un-deformed (rectilinear) bundle around its symmetry axis, and then
we bend it exactly in the same configuration as before, the
resulting relative sliding of the same pair is much smaller. The
larger the twist, the smaller the slide, Fig.8. The explanation is
simple, the twisted pair follows alternatively segments inside the
bending (positive sliding) and outside the bending (negative
sliding) and the total final slide is compensated to zero almost.
The sliding at the end of the pair is actually produced only by the
last unfinished turn of twist around the bundle axis. This result
was verified with a powerful numerical code [36] and the same
behavior was confirmed. This fact may provide insight about the
induced twist in flagellum around the body of a trypanosome-like
cell, or about the propagation of helical flagellar beats. Also, the
combination between bending and twisting could work like a molecular
gear shift: for the same bending deformation, more twist reduces the
amount of sliding (but increases torque).

The generalized sliding equation Eq.(18) is useful in a variational
formulation for the shape problem, and hence the calculated sliding
distribution can be related to molecular motors activity and
distribution for different optimal natural configurations. From
differential geometry [29], we can express ${\vec \omega}$ as a
functional of the filament unit tangent vector and its derivatives.
Consequently the sliding functional Eq.(18) can be expressed in
terms of derivatives of the filament equation only $\delta[{\vec
r}',{\vec r}'',{\vec r}''']$ which, under the request of minimum
slide with constant arc length, provides an Euler-Lagrange minimal
action system
\begin{equation}
{{\partial {\vec \Omega}}\over{\partial {\hat \omega}}} \biggl ( -
{{\partial {\hat \omega}}\over{\partial {\vec r}'}}{\vec r}' +
{{\partial {\hat \omega}}\over{\partial {\vec r}''}}{\vec r}''+
{{\partial {\hat \omega}}\over{\partial {\vec r}'''}}{\vec r}'''
\biggr ) + {\vec \Omega} =0,
\end{equation}
where the dot product between terms is performed in the tensor
contraction sense, and where ${\vec \Omega}[{\hat \omega}[{\vec
r}',{\vec r}'',{\vec r}''']]= {{d^2}\over{ds^2}}e^{\int_{0}^{s}{\hat
\omega}ds'}\cdot {\vec r}_0$. This partial differential system is
difficult to solve since $[{\vec \omega},{{d{\vec
\omega}}\over{ds}}]\neq 0$ and the derivatives of the exponential
matrix (and consequently its formal Taylor series) cannot be
expressed in a concise form, but we conjecture that the helix
provides a solution, by minimizing the total sliding.

\subsection{2-d applications}
\label{sec:52}

The 3-d sliding equation Eq.(18) can be reduce to a simpler
expression for 2-dimensional bending, by choosing $\tau=0,{\vec
b}=cst.$, and $\Lambda=0$. The 2-dimensional version is also in
agreement with the expressions of the 2-d bend for constant normal
separation from the literature ($\Delta =a$), that is $\delta=a \int
k ds $ [5,23,26,34,35]. Like Eqs.(16) in the 3-d case, we can also
obtain a nonlinear partial differential system of equations in
curvature and in the two deformations for 2-d
\begin{equation}
{{\partial ^2 \Delta}\over{\partial s^2}}+{{\partial
k}\over{\partial s}}\delta+k^2 \Delta-{{\partial k}\over{\partial
\beta}}=0, \ \ \ {{\partial \delta}\over{\partial s}}=k \Delta.
\end{equation}
As an application of the {\it inverse}problem for a 2-d system, we
investigate planar waves propagating along the bundle that provide a
traveling wave profile for the curvature. In Fig.9 we present the
waveform of the shape and curvature $g$, which induce periodic
variation of the sliding $\delta$ and transverse shift $\Delta$. We
note that the largest sliding and transverse separation occur when
the curvature has its fastest variation along the bundle. In the
right frame of Fig.9 we present a similar analysis, but in the case
of a kink soliton traveling along the bundle. The sliding, the
separation, and the curvature change in synchronism, and they are
maximum when the soliton has the fastest variation in the shape.

Conversely, one can use Eq.(20) for the {\it direct} problem.
predict the shape of the bundle for a given sliding function. This
would represent the problem of predicting waves and beats from
hypotheses over the action and synchronism of motors. For example,
we provide a sliding $\delta$ as a traveling kink, Fig.10, that is a
localized wave that flips over the state of motors.  Both curvature
and normal separation result in a localized shapes. In the bottom of
Fig. 10, we present the resulting 2-d shape of the bundle, by
integration of Eq.(20) and of the intrinsic differential equation of
the curvature [29], which strongly resembles a ciliary beat pattern.

\section{Conclusions}
\label{sec:6}

We describe the dynamics of axoneme-like filament bundles from the
perspective of the relative sliding between filament pairs, in an
arbitrary shape bundle, in order to understand and explain
mechanisms of self-deformation. The equations obtained can be used
in further modeling approaches and experimental studies in order to
calculate force and torque distributions. The analysis of filament
bundles could be placed in between differential curves and surface
theories. This study reports six expressions for sliding:  for
bending in parallel or arbitrary geometry,Eqs.(3,5)  for twisting
Eqs.(7,10), and  for arbitrary sliding involving simultaneous
bending and twisting Eqs.(11,18). We begin by analyzing two basic
types of bundle deformations, circular bending and uniform twisting.
Then we generalize the bending sliding formula for arbitrary shape,
by integration of the uniform sliding with variable parameters. We
exemplify with distributions of sliding between filaments for planar
waves, rigid rotation and helical motion. Similarly, we generalize
the sliding produced by arbitrary twisting, but in this case the
equations request numerical analysis. We obtain the formula for
uniform twisting, and for twisting around a deformed shape, and we
illustrate these formulas with numerical examples from helical, and
trypanosome-like shapes. We also analyze the influence of
non-parallelism between filaments for both bending and twisting. By
using a functional differential formalism, we obtain general formula
for the sliding produced by any 3-d shape, Eq.(18). This result can
provide the sliding distribution in any segment of any pair, for a
bundle of an arbitrary geometry, stationary or in motion. Based on
the theory of motion of curves, we obtain a differential equation
which connects the curvature, the tangent sliding, and the
transverse shift of the filaments. Based on this equation, we can
predict the distribution of the sliding (and possibly of motor
activity) for any given flagellar shape and motion. Conversely,
through these equations we can predict shapes for any arbitrary
distribution of motor activity, if this distribution can be related
with the sliding, twisting and transverse shifting. We observed that
the coupling of twisting and bending, significantly reduces the
relative sliding.

Several applications of the obtained sliding expressions are in good
agreement with present geometrical models [14,15,38,39], and
dynamical models [2,3,17,23,26,35]. The results are also discussed
form the global geometrical invariants point of view, like total
integral curvature and length, and are compared to the Hamiltonian
models of similar systems [5,20,28,34,40], as well as with nonlinear
effects observed or predicted in the literature. The equations
obtained for the distribution of sliding along and among the pairs
can be used in further modeling, theoretical approaches and/or
experimental studies in order to calculate force and torque
distributions, and--starting form experimental observed shapes--to
calculate and predict energy transfer between motors and pairs, and
in general to analyze the dynamics of the flagellum. These
conclusions expand are understanding of cellular self-propulsion
using flagellum or hair-like systems, and enhance the analysis of
bending and twisting waves that can generate swimming.

\vskip 1cm

{\it Acknowledgements} We gratefully acknowledge the support of the
Louisiana Board of Regents under grant LEQSF $(2004-2007)-RD-A25$
and are grateful for the IDEAS program
($www.scitech.nsula.edu/IDEAS$) support. A.L. gratefully
acknowledges the support of the National Science Foundation under
grant 0140274.

\vfill
\eject

{\bf Figure Captions}

\begin{enumerate}

\item
Fig.1.
Sliding induced by the bending of nine equal length $L$
filaments distributed on a base circle around a central filament.
The bending center is $O_c$, and the bending plane makes $\phi_0$
angle with $Ox$ axis. $O'$ and $R'$ are the bending center and
radius of an outer filament.

\item
Fig.2.
Density plot of the sliding $\delta_{tan}$ versus arc length
$s$ for a deformed bundle of 9 parallel filaments. From white to
black, the sliding values range from maximum negative value to its
maximum positive value, respectively. Upper left: Right helix. When
the helix rotates the density waves travel along the $s$-direction.
Upper right: Left helix. Lower left: 2-d sine wave, the pattern
travel to the right. Lower right: Rotation of a rigid bundle around
a point. The pattern in travels along the vertical axis.

\item
Fig.3.
Sliding values for five (out of nine) filaments in a circular
bend versus the arc length  $s$ of the bend. Continuous straight
lines represent the total sliding (from 0 to the current $s$) if the
filaments are parallel. The dotted lines represent the total sliding
(0 to current $s$) of the same filaments in a non-parallel
configuration. Thin oscillating curves represent the infinitesimal
sliding at the corresponding $s$. Upper left frame (3a): tangential
deviation from parallelism given by an oscillatory angular
separation between filaments. Upper right (3b): uniformly increasing
tangential deviation from parallelism. Lower left (3c): radial
oscillating non-parallelism. Lower right (3d): both tangential and
radial deviations from parallelism.

\item
Fig.4. The thin continuous lines represent infinitesimal sliding of
the 5 filaments in a parallel bundle with variable radius and
variable pitch (bundle represented in top left corner of frame)
versus $s$. The dotted lines show same sliding for coupling two
types of non-parallelism: oscillations in radius and in angular
separation.

\item
Fig.5. The relative sliding  between three adjacent filaments
subjected to constant tangential deviation from parallelism in a
circular bend, versus $s$. The bending direction is vertical
downwards in the two circular sections. The sliding
$\delta_{12},\delta_{23}$ in the case of parallel filaments, are
plotted with continuous line, while the sliding for tangentially
oscillating filaments ($\delta_{12\nshortparallel
},\delta_{23\nshortparallel }$)) is plotted with dotted line.

\item
Fig.6. Uniform twisting deformation for a circular cylinder.

\item
Fig.7. Geometry of relative sliding for a pair of filaments (1,2) in
a general 3-d deformation of the bundle. The origins of the two
tangents represent the points of same length $L$ along each of the
filaments, and these points are joined by the ${\vec dr}$ vector.
The intersection of the normal plane at $s_1 =L$ with filament 2
determines a point $s^*$ related to the sliding, $\delta=L-s^*$
(drawn with thicker line)

\item
Fig.8. Maximum sliding between two neighbor filaments ($r=2\mu,
\delta \phi =2 \pi /14, \phi =0$) plotted versus arc length. The
straight line is for a bend of radius $R=6\mu$ at $\phi_0 =\pi/2$,
Eq.(5). If the bundle is first twisted around its axis with $2\pi,
4\pi, 8\pi$ or $16 \pi$ rad, and then bend with the same radius, we
have different sliding: much smaller and oscillating. Smaller
sliding is produced by larger twisting angle. The number of periods
of sliding oscillation is given by the torsion angle divided by
$2\pi$.

\item
Fig.9. Traveling waves along the bundle. Top: Curvature $k$, tangent
sliding $\delta$ and filament separation in the normal direction
$\Delta$ versus the the arc length. Bottom: the shape of the bundle.

\item
Fig.10. A soliton ($sech^2$ profile) in curvature travels along the
bundle and produces a kink-soliton ($\tanh$ profile) in the sliding
$\delta$ distribution, a soliton ($sech^2$) in the separation shift
$\Delta$, and a symmetric pattern of 2-d beats in the bundle.

\end{enumerate}

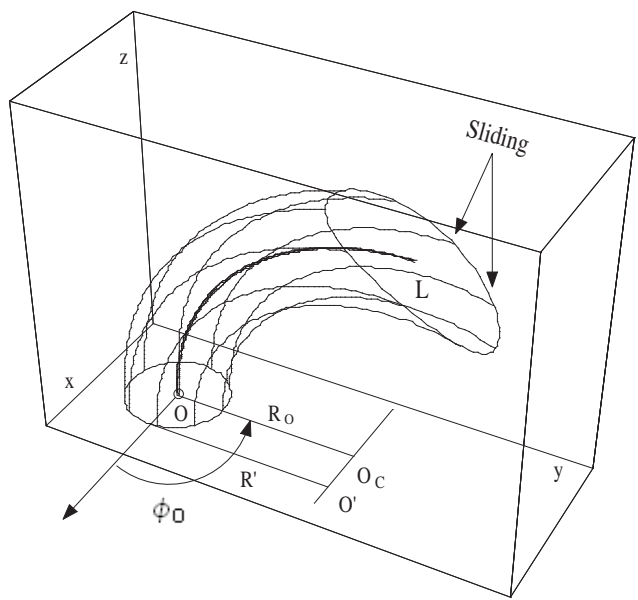
\end{document}